\documentstyle[aps,pre]{revtex}
\begin{document}
\author{B. Fourcade}
\address{Institut Laue Langevin, and \\ Maison des Magist\`eres Jean Perrin,
\\C.N.R.S., B.P.166, 25 avenue des Martyrs, \\38042 Grenoble, Cedex
9, France}
\author{A.-M.S. Tremblay}
\address{D\'epartement de physique and Centre de recherche en physique du
solide\\
Universit\'e de Sherbrooke, Sherbrooke, Qu\'ebec, J1K 2R1, Canada}
\title{Field theory and second renormalization group for multifractals in
percolation }
\date{8 September 1994, cond-mat/9409097}
\maketitle

\begin{abstract}
The field-theory for multifractals in percolation is reformulated in such a
way that multifractal exponents clearly appear as eigenvalues of a second
renormalization group. The first renormalization group describes geometrical
properties of percolation clusters, while the second-one describes
electrical properties, including noise cumulants. In this context,
multifractal exponents are associated with symmetry-breaking fields in
replica space. This provides an explanation for their observability. It is
suggested that multifractal exponents are ''dominant'' instead of
''relevant'' since there exists an arbitrary scale factor which can change
their sign from positive to negative without changing the Physics of the
problem.
\end{abstract}
\pacs{64.60.Ak, 05.70.Jk, 72.70.+m, 05.40.+j}

\section{Introduction}

The renormalization group and critical phenomena have provided, over the
years, the key concepts which allow us to understand problems ranging from
phase transitions to percolation. Even though infinite sets of exponents,
such as crossover exponents, were calculated early after the introduction of
the renormalization group in critical phenomena,\cite{mf} attention is
usually focused on a few relevant exponents. This focus on a few exponents
is justified since observable quantities in general couple to many
renormalization group eigen-operators, including the most relevant ones,
which eventually dominate their behavior. That lore was challenged
relatively recently by the appearance of infinite sets of measurable
exponents in various fields. Problems where such infinite sets occur are
collectively known as multifractal problems,\cite{hasley}\cite{grant}\cite
{paladin} even though some of these problems have a quite different physical
nature. We discuss here the problem of electrical properties of percolating
networks.\cite{amst}\cite{arcangelis}\cite{review}

The difficulties associated with the formulation of a Lagrangian field
theory for multifractals have been first discussed by Ludwig and Duplantier.%
\cite{dl}\cite{deutsch}. In particular, the exponents $x_n$ describing
multifractal behavior are a convex function of $n$, while the analogous
exponents of field powers in a field theory are in general concave, as a
consequence of stability and correlation inequalities.\cite{dl} But, as
stated in Ref. \cite{dl}, powers of composite operators, like derivatives,
can exhibit multifractal behavior. It is therefore possible to formulate for
multifractals a field theory which is amenable to a renormalization group
analysis. This has been done for percolation by Parks, Harris and Lubensky%
\cite{phl} (PHL). This field theory however does have some peculiarities
which make it different from usual field theories. In the present paper, we
reinterpret the field theory of PHL so as to make special features clearer.
In fact there are also some differences in the way we set up and interpret
the field theory. A detailed discussion of the differences and of the
reasons which motivate our approach have appeared.\cite{bibi} The structure
which will emerge here is that geometrical properties of percolation
clusters are described by a standard field theory on which one can perform a
renormalization group analysis. We call this the ''first renormalization
group''. By contrast, multifractal properties, originating from the
electrical properties of the cluster, are described by a ''second
renormalization group''.\cite{cercle}\cite{rg2} The structure of this second
renormalization group depends on the first renormalization group: Once the
usual recursion relations for percolation are derived, the recursion
relation for each multifractal moment must be found by a further projection
onto an appropriate eigenbasis. In the second renormalization group, the
role of the fields is played by the microscopic noise cumulants $v_s$. These
fields $v_s$ are conjugate to replica-space gradients of the operators, for
example ${\bf k}^{2s}\Phi _{{\bf k}}({\bf q)}\Phi _{-{\bf k}}(-{\bf q)}$
where ${\bf q}$ is defined in the usual Brillouin zone while ${\bf k}$ is
the replica-space extension of the Fourier variable for the voltage. Without
changing the Physics, the scale of ${\bf k}$ can be changed by an arbitrary
factor. The existence of this arbitrary scale factor is a direct consequence
of the linearity of Kirchhoff's laws. This has no equivalent in usual
critical phenomena. Due to this arbitrary scale, the multifractal exponents
can change sign, rendering inappropriate the usual classification as
relevant or irrelevant. We suggest, therefore, to call the multifractal
exponents ''dominant'' exponents since they determine the leading scaling
behavior of observable quantities while corrections would come from
''sub-dominant'' exponents. The fields for multifractal moments are
associated with operators which break rotational (permutation\cite{dl})
symmetry in replica space. This seems to be the reason why the multifractal
moments are each associated with a different dominant exponent, and not just
to a few ones.

In the following section, we make a short review of the phenomenology of
multifractals in percolation. Then we derive the field theory and proceed
with the renormalization groups. The discussion section clarifies some
points of interpretation.

\section{Phenomenology of multifractals in percolation}

For completeness, let us recall how the infinite set of exponents appears in
percolation through the problem of noise. Suppose that the conducting
resistors of a percolating network are fluctuating independently in time.
The total resistance of a given network is then a random variable in time
whose cumulants depend on those of each component resistor. The cumulants of
a given order are assumed to be the same for all component resistors. The
cumulants of the total resistance $R$ are, in principle, accessible
experimentally, and measurements of the second cumulant, corresponding to $%
1/f$ noise, have actually been performed.\cite{garfunkel} Theoretically, for
systems of finite size $L$ at bulk criticality, one finds, after averaging
over the microscopic noise, that the cumulants,$\,C_R^{\left( n\right)
}\left( L\right) $, of order $n$ scale as
\begin{equation}
\label{micum}<C_R^{\left( n\right) }\left( L\right) >_C=v_n<\sum_\alpha
i_\alpha ^{2n}>_C\sim L^{-x_n}
\end{equation}
where $C$ represents average over percolating lattice configurations, $%
i_\alpha $ is the current that flows in branch $\alpha $ of the time
averaged network and where $v_n$ is the amplitude of the n'th cumulant of
the elementary resistance fluctuations. For example, the usual electrical
noise amplitude for one microscopic resistance $r$ is obtained from
\begin{equation}
\label{v2}v_2=\left\{ \delta r\delta r\right\} _f
\end{equation}
where $\left\{ {}\right\} _f$ refers to time average over the noise. The
first equality in Eq. (\ref{micum}) follows from Cohn's theorem
\begin{equation}
\label{cohn}\delta R=\sum_\alpha \delta r_\alpha \,i_\alpha ^2
\end{equation}
where the total input current, $I_{inj.}$, is unity. Each exponent $x_n$ is
different and is not a simple linear function of $n$, as commonly occurs in
critical phenomena under the name ''gap scaling''.\cite{mf} Such an infinite
set of exponents also arises in the other analogous problems mentioned
above. In all these problems, one is assigning to parts of a fractal network
a weight, or a measure, which is obtained from the solution of Laplace's
equation. Here that weight corresponds to the power dissipated in a bond.

Finally, let us recall that the positive integer moments, $<\sum_\alpha
i_\alpha ^{2n}>_C$, suffice to characterize completely the distribution of
the currents flowing through the network.\cite{pierre} But, as stated
previously, the {\it exact} value of the integer moments is necessary to
reconstruct all the information so that the leading scaling behavior of the
positive moments does not suffice to find, for example, negative moments. In
the following, we concentrate only on the scaling behavior of the positive
integer moments. Note that in order to keep the same notation as PHL\cite
{phl}, we work with exponents defined by
\begin{equation}
\psi _s/\nu \equiv -x_n\quad \quad (s=n).
\end{equation}

\section{Field theory for the generating function}

\subsection{Generating function for the cumulants}

Since Kirchhoff's law can be obtained by minimizing the entropy production,
or the dissipated power in the electrical network, it is natural to start
from a generating function for the resistance between two points $y$ and $%
y^{\prime }$ of the network
\begin{equation}
\label{generating}W\left( k;y,y^{\prime }\right) =\int_{\Delta V_{\min
}}^{\Delta V_{\max }}d\left[ V(x)\right] \exp \left( -H+ik\,\left[
V(y)-V(y^{\prime })\right] \right)
\end{equation}
where $\Delta V_{\min }$ and $\Delta V_{\max }$ are, respectively, the
minimum and maximum voltage drops for a finite size system, and where $H$ is
given by,
\begin{equation}
\label{hamiltonien}H\equiv \frac 12\sum\limits_{<x,x^{\prime }>}\,\sigma
_b(x,x^{\prime };C,f)\,\left[ V(x)-V(x^{\prime })\right] ^2
\end{equation}
with $\sigma _b(x,x^{\prime };C,f)$ the conductivity of the bonds $\alpha $
linking each pair of points $x$ and $x^{\prime }$ for a given configuration $%
C\ $of the random resistor network. For each given configuration of the
random network, the conductivities fluctuate in time. In other words,
\begin{equation}
\label{microfluct}\sigma _b(x,x^{\prime };C,f)=\sigma _0(x,x^{\prime
};C)(1+\epsilon ),
\end{equation}
where $\epsilon \ll 1$ is a random variable whose probability distribution
is given by $f$ $\left( \epsilon \right) $. Hence, there are two types of
averages to perform: The usual bond-disorder average and, for each lattice
configuration, an average over the microscopic noise. To obtain a
Hamiltonian with the same structure as for a spin system, (e.g. $J\left(
x-x^{^{\prime }}\right) S(x)S(x^{^{\prime }})$) one uses Fourier transforms.
Since the potential differences are on a bounded interval, determined by the
boundary conditions, it is possible to use discrete Fourier series.
Formally, we may write
\begin{equation}
\label{fourier}H=\sum\limits_{<x,x^{\prime }>}\,\sigma _b(x,x^{\prime
};C,f)\sum\limits_kA_ke^{ik(V(x)-V(x^{\prime }))}
\end{equation}
where the Fourier coefficients are given by
\begin{equation}
\label{definea}A_k={\frac 1{\Delta V_{\max \,}}}\,\int_{\Delta V_{\min
}}^{\Delta V_{\max }}\,d(\Delta V)\,\left[ \Delta V\right] ^2e^{ik\Delta V}.
\end{equation}
Because the $k=0$ terms correspond to a uniform distribution of voltage,
they will be discarded. In the theory of Stephen,\cite{stephen} the order
parameter $\psi $ is the defined in such a way that its autocorrelation
function vanishes in the non-percolating phase and decays exponentially in
the percolating phase. This order parameter is
\begin{equation}
\label{orderpara}\psi _k(y)\equiv e^{ikV(y)}
\end{equation}
\thinspace where the wave vector $k$ takes a discrete number of values $2\pi
n/(\Delta V_{\max }).$ The presence of a lattice, instead of a continuum,
leads to the existence of an ultraviolet cutoff $\Lambda _k$ corresponding
to the minimum voltage\cite{fasterthanL} $\Lambda _k=2\pi /(\Delta V_{\min
}) $. That cutoff becomes infinite in the percolation limit $\sigma
_o^{-1}\rightarrow 0$. For a given realization, we may then write $H$ in the
form advertized, namely
\begin{equation}
\label{exphamil}H\,\,=\sum\limits_kA_{k\,}\,\sum\limits_{<x,x^{^{\prime
}}>}\sigma _b(x,x^{\prime };C,f)\,\psi _k(x)\,\psi _k(x^{\prime })\,
\end{equation}
where, in the second sum, we consider only the connected links of the
network. To understand what follows, it is important to realize that the $k$%
's scale as $\Delta V_{\max }$ $^{-1}$ and that we must use a discrete
Fourier expansion, since the replica method is only valid for finite system.
The infinite-system limit is taken at the very end.

In the limit where $\sigma _0^{-1}\rightarrow 0$, the saddle point
configuration for the voltage drops obeys Kirchhoff's laws and, in this
limit, we obtain for the generating function
\begin{equation}
\label{saddle}W\left( k;y,y^{\prime }\right) =<\psi _k(y)\,\psi
_{-k}(y^{\prime })>\,=\,\;Z\;e^{-\frac 12k^2R(y,y^{\prime };C,f)}
\end{equation}
where
\begin{equation}
\label{4.9}Z\,=\int \,d\left[ V(x)\right] \,\,e^{-\frac 12%
\,\sum\limits_{<x,x^{\prime }>}\,\sigma _b(x,x^{\prime };C,f)\,\left[
V(x)-V(x^{\prime })\right] ^2}=\int \,d\left[ V(x)\right] \exp \left(
-H\right)
\end{equation}
\thinspace and $R(y,y^{\prime };C,f)$ is the resistance between nodes $y$
and $y^{\prime }$ for a given configuration $C$ of the random network, and
realization $f$ of the noise. As in PHL \cite{phl} the cumulant averages for
the resistance noise may be obtained from the generating function. We first
average Eq. (\ref{4.9}) over the probability distribution for the noise $f$
and then we average over the disorder as follows,
\begin{equation}
\label{4.11a}\left\langle \left\{ \exp \left( -\frac 12k^2\,R\left(
y,y^{\prime };C)\right) \right) \right\} _f\,\,\nu \left( y,y^{\prime
};C\right) \right\rangle _C
\end{equation}
where the function $\nu $ plays the role of a conditional probability that
the two points $y$ and $y^{\prime }$ are connected. Expanding the left-hand
side in cumulants of the resistance, Eq. (\ref{4.11a}) becomes,
\begin{equation}
\label{4.11}\left\langle \exp \left[ \sum_{s\geq 1}\frac{\left( -1\right) ^s%
}{2^s\,s!}\,k^{2s}\,C_R^{\left( s\right) }\left( y,y^{\prime };C)\right)
\right] \nu \left( y,y^{\prime };C\right) \right\rangle _C=\left\langle
\left\{ \left\langle \psi _k\left( y\right) \psi _{-k}\left( y^{\prime
}\right) \right\rangle \right\} _f\right\rangle _C.
\end{equation}

The cumulants of the resistance fluctuations $C_R^{(s)}(y,y^{\prime };C)$
may thus be obtained from derivatives of the generating function since, by
Eq.(\ref{micum}) and in the $\sigma _o^{-1}\rightarrow 0$ limit, they are
proportional to the microscopic cumulants,\cite{amst}
\begin{equation}
\label{4.12}
\begin{array}{c}
\left\langle C_R^{\left( s\right) }\left( y,y^{\prime };C)\right) \nu \left(
y,y^{\prime };C\right) \right\rangle _C= \\
\,\left( -1\right) ^s2^s\,s!\,k^{-2s}\,v_s\frac{\partial \left[ \left\langle
\left\{ \left\langle \psi _k\left( y\right) \psi _{-k}\left( y^{\prime
}\right) \right\rangle \right\} _f\right\rangle _C\right] }{\partial \,v_s}%
\mid _{v_l=0\,,\,\,\forall \,l}.
\end{array}
\end{equation}
Thus, the macroscopic cumulants of the noisy resistor network can be
obtained from the autocorrelation of the order parameter, after averaging
over bond disorder. It is important to note that the quantities $v_s$ play
the role of fields whose conjugate operators will contain polynomials in $k$%
, as explained later.

\subsection{Replica method and effective action}

Since there are two types of averages over random variables, we introduce
two types of replica, as suggested by PHL \cite{phl}: $N$ replica for the
average over noise $(f)$, and $M$ replicas for the average over bond
disorder $(C)$:
\begin{equation}
\label{4.14}\left\langle \left\{ LogZ\right\} _f\right\rangle _C=\lim
_{N\rightarrow 0\\M\rightarrow 0}\,\frac{\left( \left\langle \left[ \left\{
Z^N\right\} _f\right] ^M\right\rangle _C\,-1\right) }{N\,M}
\end{equation}
As usual, the limiting process is justified only for finite systems. In
other words, the limit $L\rightarrow \infty $ is taken after the limits $N$,
$M\rightarrow 0$.

We do not repeat the details of the derivation of the field theory.\cite{phl}
As usual, it proceeds by introducing Hubbard-Stratonovich variables $\Phi $
conjugate to each $\psi $ appearing in $Z^{NM}$. Expanding in powers of $%
\psi $ and performing the integrals over the voltages appearing in the
original generating function, one generates a power series in $\Phi $ which
can be re-exponentiated to yield an effective action for the $\Phi $. These
variables are now the {\it operators} of the field theory. For
Hubbard-Stratonovich transformations, the generating function $<\psi _{{\bf k%
}}(x)\psi _{{\bf -k}}(x)>$ is simply proportional to $<\Phi _{{\bf k}}\left(
{\bf x}\right) \Phi _{-{\bf k}}\left( {\bf x}\right) >$, hence all we need
is the effective action for the operators $\Phi _{{\bf k}}$%
\begin{equation}
\label{4.15}
\begin{array}{c}
{\cal L}\left( \Phi \right) =\frac 12\int {\it d}^dx\,\Delta k\sum_{{\bf k}%
}r_{{\bf k}}\left[ \Phi _{{\bf k}}\left( {\bf x}\right) \Phi _{-{\bf k}%
}\left( {\bf x}\right) +\nabla \Phi _{{\bf k}}\left( {\bf x}\right) \cdot
\nabla \Phi _{-{\bf k}}\left( {\bf x}\right) \right] \\ +\frac{u_3}6\,\int
{\it d}^dx\left( \Delta k\right) \,^2\sum_{{\bf k}_1+{\bf k}_2\neq 0}\Phi _{%
{\bf k}_1}\left( {\bf x}\right) \Phi _{{\bf k}_2}\left( {\bf x}\right) \Phi
_{-{\bf k}_1-{\bf k}_2}\left( {\bf x}\right) .
\end{array}
\end{equation}

The ${\bf k}$ in Eq. (\ref{4.15}) is any of the ${\bf k}$ Fourier variables
whose components are labeled $k_{\alpha \beta }$ in the $NM-$ dimensional
replica space and ${\bf k^2\equiv }\sum_{\alpha \beta }k_{\alpha \beta }^2$
is the square-modulus of ${\bf k}$. There are $(L^{MN}-1)$ operators $\Phi _{%
{\bf k}}(x)$ at spatial point $x$ (the ${\bf k}=0$ case is omitted). The $%
k_{\alpha \beta }$ are conjugate to the electrical potentials of the
replicated systems. They contain a scale factor correponding to the scale of
the electrical potential. In the limit of geometrical percolation (i.e. no
transport property), the $k_{\alpha \beta }$ are all zero. Note that despite
the notation, the ${\bf k}${\bf \ }are tensors of rank one and not two, as
far as ''rotations'' in replica space are concerned. As usual, to find
critical exponents the ${\bf k}$ dependence of the coupling constant $u_3$
can be neglected, but that of $r_{{\bf k}}$ is crucial. The scaling of the
terms entering $r_{{\bf k}}$ can be infered from the scaling properties of
the generating function. Writing explicitly the dependence on the
microscopic cumulants $v_s$, one obtains for the scaling properties of the
generating function,
\begin{equation}
\label{scalingG-vs}G_{{\bf k}}\left( y-y^{\prime },p-p_c,\left\{ v_s\right\}
_{s\geq 1},I_{inj.}\right) =G_{\lambda \,{\bf k}}\left( y-y^{\prime
},p-p_c,\left\{ \lambda ^{-2s}v_s\right\} _{s\geq 1},I_{inj.}\right)
\end{equation}
since ${\bf k}$ scales as ${\frac 1{\Delta V_{max}}}$. Indeed, the cumulants
$C_R^{(s)}(y,y^{\prime };C)$ are linearly proportional to the microscopic
cumulants $v_s$ so that, as can be seen from Eq. (\ref{4.11}), the
autocorrelation function for the order parameter does have the scaling
property (\ref{scalingG-vs}). Expanding for small $v_s$, the most general
form for the $r_{{\bf k\,}}$ is then an homogeneous polynomial in ${\bf k}$
and $v_s$.

\begin{equation}
\label{4.28}r_{{\bf k\,}}=p-p_c+\sum_{s\geq 1}{\bf k}^{2s}\left(
\sum_{s_1+s_2+...+s_i=s}v_{s_1}v_{s_2}...v_{s_i}\,P_{(s_1,s_2,...,s_i)}%
\left( ...,\theta _{{\bf k}_{\alpha ,\beta }},...\right) \right) .
\end{equation}
In Eq.(\ref{4.28}) the functions $P_{{\bf s}}\left( ...,\theta _{{\bf k}%
},...\right) $ depend on the angular variables $\theta _k$ in the $n=MN$
dimensional replica space. For each ${\bf s=}(s_1,s_2,...,s_i)$, these
polynomials can be expressed as a linear combination of the spherical
harmonics for the $MN$ angular variables $\theta _{{\bf k}}$ . Their
explicit expressions depends on the particular distribution $f$ for the
noise of the elementary bond resistor. The expansion in Eq. (\ref{4.28}) is
justified by the fact that for a finite system, the effective action is an
analytic function of ${\bf k}$ and hence, by scaling arguments, of $v_s.$
For $s=1$, it will be seen in the Appendix that $P_1\left( \theta _{{\bf k}%
}\right) =1$, {\it i.e.} ${\bf k}^2$ is the only polynomial of degree 2
which contributes and it is an eigenpolynomial of the second renormalization
group.

Finally, observable quantities are obtained using the standard
replica-method identity,
\begin{equation}
\label{a425}
\begin{array}{c}
G_k\left( y-y^{\prime },p-p_c,\left\{ v_s\right\} _{s\geq 1},I_{inj.}\right)
=\left\langle \left\{ \left\langle \psi _k(y)\psi _{-k}(y^{\prime
})\right\rangle \right\} _f\right\rangle _C \\
\sim \lim _{N\rightarrow 0\,M\rightarrow 0}\left\langle \Phi _{{\bf k}%
}(y)\Phi _{-{\bf k}}(y^{\prime })\right\rangle
\end{array}
\end{equation}
In the absence of symmetry breaking in replica space, the modulus of the
replicated ${\bf k}$ in Eq. (\ref{a425}) is equal to $k$ in expression (\ref
{4.11}).

\subsection{Scaling properties}

Near the percolation critical point, the generating function scales as,
\begin{equation}
\label{4.29}
\begin{array}{c}
\lim _{L\rightarrow \infty \,;\,\,\sigma _0^{-1}\rightarrow
0\,;\,\,n\rightarrow 0}G_k\left( y-y^{\prime },p-p_c,\left\{ v_s\right\}
_{s\geq 1},I_{inj.}\right) = \\
\lim _{L\rightarrow \infty \,\,;\,\,\sigma _0^{-1}\rightarrow
0\,\,;\,\,n\rightarrow 0}\lambda ^{2-\eta _p-d}G_k\left( \left( y-y^{\prime
}\right) \lambda ^{-1},\left( p-p_c\right) \lambda ^{1/\nu _p},\left\{
v_s\,\lambda ^{\psi _s/\nu _p}\right\} _{s\geq 1},I_{inj.}\right)
\end{array}
\end{equation}
which also implies
\begin{equation}
\label{4.30}
\begin{array}{c}
\lim _{L\rightarrow \infty \,\,;\,\,\sigma _0^{-1}\rightarrow
0\,\,;\,\,n\rightarrow 0}G_k\left( y-y^{\prime },p-p_c,\left\{ v_s\right\}
_{s\geq 1},I_{inj.}\right) = \\
\left( p-p_c\right) ^{\left( d-2+\eta _p\right) \nu _p}S\left( \left(
y-y^{\prime }\right) \left( p-p_c\right) ^{\nu _p},\left\{
v_s\,(p-p_c)^{-\psi _s}\right\} _{s\geq 1}\right)
\end{array}
\end{equation}
where $S$ is a scaling function. As usual, the correlation length $\xi $
behaves as $\left( p-p_c\right) ^{-\nu }$. The $L\rightarrow \infty $ must
be taken at the end for the replica method to be valid. From the way the
problem is set up, the $n\rightarrow 0$ limit must be taken before the $%
\sigma _0^{-1}\rightarrow 0$ limit. However, as we shall see in the
following sections, we need to add another condition to linearize the second
renormalization group equations, namely that
$$
v_s<<v_1\quad ;\quad \forall s>1
$$
This inequality is consistent with the phenomenology of $1/f$ noise.

\section{Renormalization group approach:}

\subsection{First, and second renormalization group.}

The renormalization group equations are obtained\cite{phl} from the usual
procedure for a cubic Landau-Ginzburg functional. Using the Wilson approach,
operators $\Phi _{{\bf k}}({\bf q})$ whose wavevector ${\bf q}$ is in a
shell $\Lambda /b<q\leq \Lambda $, near the cut-off $\Lambda $ coming from
the physical lattice, are traced over. It is always possible to choose the
lattice spacing units such that $\Lambda =1$. Lengths are then rescaled by $%
b $, while the operators are rescaled as follows:
\begin{equation}
\label{recur1}\Phi _{{\bf k}}^{^{\prime }}({\bf q^{\prime }}=b{\bf q}%
)=b^{(-d-2+\eta _p)/2}\Phi _{{\bf k}}({\bf q})
\end{equation}
Let $K_d$ be the surface of the $d-$dimensional sphere. To one-loop order,
in dimension $d=6-\epsilon $ near the upper-critical dimension $d=6$, the
differential recursion relations for $r_{{\bf k}}$ and for the coupling
constant $g=K_du_3^2/2$ take the form\cite{phl}\cite{hl},
\begin{equation}
\label{4.19}\frac{dr_{{\bf k}}}{dl}=\,\left( 2-\eta _p\right) r_{{\bf k}%
\,}-g\Sigma _{{\bf k}}
\end{equation}
$$
\frac{dg}{dl}=\left( \epsilon -3\eta _p\right) g-8g^2
$$
where $l$ is defined by $b=e^l$. In the first equation, $\Sigma _{{\bf k}}$
is the self-energy-correction to one loop \cite{phl}
\begin{equation}
\label{A.11}
\begin{array}{c}
\Sigma _{
{\bf k}}=\lim _{n\rightarrow 0}(\Delta k)\sum_{{\bf p,p+k\neq 0}}\,G\left[
{\bf p}\right] G\left[ {\bf p+k}\right] . \\ \equiv -2G\left[ {\bf k}\right]
G\left[ {\bf 0}\right] +\widetilde{\Sigma }_{{\bf k}}
\end{array}
\end{equation}

Since we are looking for a linear renormalization group in $v_s$, it
suffices to restrict ourselves to the Green's functions expanded to first
order in $v_s$. Furthermore, the momentum-shell integral is for ${\bf q}%
^2=\Lambda ^2=1$ so that the Green's functions appearing in (\ref{A.11}) are
of the form,

\begin{equation}
\label{A.12}G\left[ {\bf p+k}\right] ^{-1}=1+p-p_c+\sum_{s\geq 1}v_s{\bf k}%
^{2s}P_{(s)}^{(1)}\left( ...,\theta _{{\bf k}},...\right) .
\end{equation}

The usual percolation fixed-point, $v_s=0$, and $p=p_c$ describes the
geometrical properties of percolation clusters. These properties are
completely independent of the electrical transport properties of the
network. In other words the field $r_0$ is the coefficient of a polynomial
of order $0$ in ${\bf k}$ so that the calculations of the fixed point and of
the usual geometrical exponents $\eta _{p,}\nu _p$ are independent from the
values of the fields $v_s$ which are all associated with higher-order
polynomials in ${\bf k}$ times the field operator $\Phi _{{\bf k}}\Phi _{-%
{\bf k}}$. Hence one obtains the same results as in ref.\cite{phl} namely $%
g^{*}=\epsilon /7$ for the fixed-point value of the coupling constant and $%
\eta _p=-\epsilon /21,\quad \nu _p=\frac 12+\frac{5\epsilon }{84}$ for the
exponents. These exponents, and fixed-point values of $g$ and $r_0$ come
from what we call the first renormalization group, whose predictions relate
to the geometrical properties of the percolation network, and hence are
independent of electrical properties.

Going further to obtain the exponents $\psi _s$ associated with $v_s$, one
has to linearize Eq. (\ref{4.19}) in $v_s$. To the order we are working, the
$v_s$ are all multiplied by the same power $\Phi _{{\bf k}}$, namely $\Phi _{%
{\bf k}}^2$, but by different powers of ${\bf k}$, namely a homogeneous
polynomial of order $2s$. Hence, we consider the renormalization group as a
renormalization equation in the space of homogeneous polynomials in ${\bf k}$%
. This is what we call the ''second renormalization group''.

To be more specific, let us start with the simple case of $d$ larger than
the upper critical dimension ($d_c=6$). The renormalization group equations
for $r_{{\bf k}}$ then read,
\begin{equation}
\label{abs1}\frac{dr_{{\bf k}}}{dl}=2r_{{\bf k}}.
\end{equation}
In other words, the recursion relation for the $v_s$ is obtained by
linearizing around the fixed-point $v_s=0$, $s=0,1...$ To first order in $%
v_s $, one has
\begin{equation}
\label{abs2}r_{{\bf k\,}}=p-p_c+\sum_{s\geq 1}{\bf k}^{2s}v_sP_s\left(
...,\theta _{{\bf k}_{\alpha ,\beta }},...\right)
\end{equation}
and thus the scaling in Eq.(\ref{recur1}) leads to,
\begin{equation}
\label{abs3}\frac{dv_s}{dl}=2v_s.
\end{equation}
Above six dimensions then, any $P({\bf k})={\bf k}^{2s}P_s\left( ...,\theta
_{{\bf k}_{\alpha ,\beta }},...\right) $ is an eigenvector of the second
renormalization group. The eigenvalues are all identical, as is already
known.\cite{amst} Thus, even though the precise form of $P_s({\bf k})$
depends on the microscopic noise distribution, the eigenvalue for $v_s$ does
not. Below six dimensions, things are less trivial since the self-energy$%
\widetilde{\text{ }\Sigma }_{{\bf k}}$ depends functionaly on $P_s({\bf k})$%
. Nevertheless, we will show that there exists an eigenbasis of polynomials
in ${\bf k\ }$whose eigenvalues are labeled only by the order of the
polynomial. This means that whatever the starting $P_s({\bf k})$, the
eigenvalue for $v_s$ depends only on $s$.

Below six dimensions, the RG equations must be obtained by linearizing the
self-energy $\widetilde{\Sigma }_{{\bf k}}$ as a function of $v_s\,,\,s\geq
2 $. We show in the Appendix that the eigenpolynomial basis is obtained, in
the limit $v_s<<v_1$, by solving for each integer $s$ the following equation
\begin{equation}
\label{self-Laplace}
\begin{array}{c}
-2\int_0^{+\infty }du\int_0^{+\infty }dt\,\exp \left\{ -\left( u+t\right)
\left( 1+r_0\right) \right\} \left[
\frac t{t+u}\right] ^{2s}u \\ \left\{ \exp \left( \frac{u+t}{4v_1}\,t^{-2}%
{\bf \Delta }\right) P\left( {\bf k}\right) \right\} =d_sP\left( {\bf k}%
\right)
\end{array}
\end{equation}
which correponds to the limits $\sigma _0^{-1}\rightarrow 0$ and $%
n\rightarrow 0$ for the self-energy. In Eq. (\ref{self-Laplace}) the
Laplacian operator ${\bf \Delta }$ acts on ${\bf k}$ and its limit, as $%
n\rightarrow 0$, is discussed in the next section. Because $r_{{\bf k}}$ in
Eq. (\ref{4.19}) is linear in $v_s$, the eigenvalues for the ''second
renormalization group'' are obtained as $2-\eta _p-g^{*}\left( 2+d_s\right) $%
. Everywhere below, $g$ will be taken at its fixed-point value, $g^{*}$.

\subsection{Eigenvalues and eigenvectors of the second renormalization group.
}

Eq. (\ref{4.19}) and (\ref{self-Laplace}) define the renormalization group
equation for homogeneous polynomials of order $2s$. To diagonalize this
equation we need eigenfunctions for the Lapalacian operator. In $n$%
-dimensions, the most general form for these eigenfunctions is given by \cite
{erdeleyi}
\begin{equation}
\label{A3.3}k^{-\frac{n-2}2}J_{2r+\frac{n-2}2}\left( \gamma k\right)
\,Y_{2r}\left( ...,\theta _{{\bf k}},...\right)
\end{equation}
where $Y_{2r}(...,\theta _i,...)$ is a spherical harmonic of order $2r$ in $%
n=NM$ dimensions, $J_r$ is the Bessel function of order $r$ with eigenvalue $%
-\gamma ^2$. To expand polynomials of order $2s$ on this basis, we first
recall that the maximum value of the replica-space $k\equiv \sqrt{{\bf k}^2}$
is given by the ultraviolet cutoff $\Lambda _k=k_{\max }=2\pi /\Delta
V_{\min }=\sigma _02\pi /i_{\min }$. We can then proceed as follows. First
note that an arbitrary polynomial of order $2s$ can be written as,
\begin{equation}
\label{4.38}P_s\left( k_{\alpha \beta }\right) ={\bf k}^{2s}\,\sum_{0\leq
r\leq s}\,a_{s,r}\,Y_{2r}\left( ...,\theta _{{\bf k}},...\right)
\end{equation}
where, to simplify the notation, we do not write internal indices related to
the degeneracy of the spherical harmonics. Then, we write
\begin{equation}
\label{A3.5}
\begin{array}{c}
{\bf k}^{2s}Y_{2r}\left( ...,\theta _{{\bf k}},...\right) =\left( {\bf k}%
^{2s}-{\bf k}^{2r}\Lambda _k^{2\left( s-r\right) }\right) Y_{2r}\left(
...,\theta _{{\bf k}},...\right) \\ +\,\,{\bf k}^{2r}\Lambda _k^{2\left(
s-r\right) }Y_{2r}\left( ...,\theta _{{\bf k}},...\right) .
\end{array}
\end{equation}
The last term is an eigenfunction of the Laplacian with eigenvalue zero,
while the first term can be expanded using the Fourier-Bessel expansion,\cite
{watson} which is uniformly valid for functions with zero value at the end
of the interval.
\begin{equation}
\label{A3.6}
\begin{array}{c}
\left(
{\bf k}^{2s}-{\bf k}^{2r}\Lambda _k^{2\left( s-r\right) }\right)
Y_{2r}\left( ...,\theta _{{\bf k}},...\right) = \\ \sum_{l\geq 1}b_lk^{-%
\frac{n-2}2}J_{2r+\frac{n-2}2}\left( \frac k{\Lambda _k}\zeta _l\right)
Y_{2r}\left( ...,\theta _{{\bf k}},...\right)
\end{array}
\end{equation}
where $\left\{ \zeta _\ell \right\} _{\ell \geq 1}$ are the zeros of the
Bessel function of order $2r+(n-2)/2$.

We can now substitute this expansion in the kernel of the integral appearing
in the recursion relation Eq. (\ref{self-Laplace}). The contribution of a
general term of the series will have the form,
\begin{equation}
\label{A3.7}2\int_o^\infty du\int_0^\infty dt\exp \left[ -\frac{u+t}{2v_1}%
t^{-2}\left( \frac{\zeta _l}{\Lambda _k}\right) ^2\right] k^{^{-\frac{n-2}2%
}}J_{2r+\frac{n-2}2}\left( \frac k{\Lambda _k}\zeta _l\right) Y_{2r}\left(
...,\theta _{{\bf k}},...\right)
\end{equation}
The $n=NM\rightarrow 0$ limit is obvious. Taking the $\sigma
_0^{-1}\rightarrow 0$ limit simplifies further the equation considerably
since, in the argument of the exponential, $\left[ v_1\Lambda _k^2\right]
^{-1}$ is proportional to $\sigma _0^{-1}$. This means that effectively
every term of the series behaves as if it had the same eigenvalue for the
Laplacian, namely $0$. Since $k/\Lambda _k$ is bounded between $0$ and $1$,
the series can be resummed, and any polynomial of order $2s$ is an
eigenvector with eigenvalue\cite{phl}
\begin{equation}
\label{A3.9}d_s=-2\int_0^{+\infty }du\int_0^{+\infty }dt\,\exp \left(
-\left( u+t\right) \right) \left[ \frac t{t+u}\right] ^{2s}u=-\frac 2{\left(
2s+1\right) \left( s+1\right) }.
\end{equation}

The main conclusion of the section is that the space of homogeneous
functions of order $2s$ can be characterized by a single eigenvalue which
depends only on the {\it order} of the polynomial. The form of the
microscopic distribution of the noise is, therefore, not relevant for
scaling properties of the macroscopic resistance fluctuations. Only the
condition $v_s<<v_1\,,\,\forall s>1$ which was used to derive Eq.(\ref
{self-Laplace}) needs to be satisfied. The critical exponents associated
with the fields $v_s$ are thus
\begin{equation}
\label{4.43}\psi _s=\nu _p\left( 2-\eta _p-g^{*}\left( 2+d_s\right) \right)
=1+\frac \varepsilon {7\left( s+1\right) \left( 2s+1\right) },\,s=1,2...
\end{equation}
as found by Parks, Harris and Lubensky ($\nu _p=\frac 12+\frac 5{84}\epsilon
\,\,;\,\eta _p=-\frac \epsilon {21}$).\cite{phl} It is shown in the Appendix
that this formula applies also for the case $s=1$, as written above.

\section{Discussion}

\subsection{Gap scaling}

The scaling of the usual thermodynamic observables is normally trivially
obtained from a few exponents only. This is usually refered to as ''gap
scaling''. Gap scaling also occurs in the present case. For example,
\begin{equation}
\frac{\partial ^\ell G_{{\bf k}}(x,x^{\prime })}{\partial v_s^\ell }\rfloor
_{v_s=0}\sim \left| x-x^{\prime }\right| ^{\ell \psi _s/\nu _p}
\end{equation}
That this applies to multifractals in percolation was verified by numerical
simulations in\cite{troubadours}. Clearly, one can also define universal
amplitude ratios.\cite{pierre}\cite{troubadours}

\subsection{Symmetry breaking}

With $v_s=0$ for all $s$, the action is invariant under the global
transformation
\begin{equation}
\label{12}\Phi _{{\bf k}}^{\prime }=\Phi _{{\cal R}{\bf k}}
\end{equation}
where ${\cal R}$ is a rotation (permutation) of the vector ${\bf k}$ in the
replica space of dimension $MN$. In other words, the action transforms
according to the unit representation of the group $O(MN)$. When $v_s\neq 0$,
that symmetry is broken, since polynomials of higher degree transform like
higher-dimensional representations of the group $O(MN)$.\cite{harmonique} To
have every $v_s$ associated with a different representation of the symmetry
group is a necessary but not sufficient condition to have an infinity of
observable exponents. Indeed, operators of different symmetry could couple
when higher-order corrections to the $\epsilon $ expansion are evaluated.
All this is analogous to what happens with symmetry-breaking fields in other
critical phenomena models, such as the $XY$ model for example.\cite
{fourcadexy}. The exponents $\psi _s$ here are analogous to the crossover
exponents $\varphi _n$ of the $XY$ model.\cite{xy}.

\subsection{Dominant exponents}

At first sight, the perturbations associated with the $v_s$ are all relevant
since all the exponents $\psi _s$ are found to be larger than zero. There
are two important differences however with critical phenomena (say the case
of the $XY$ model)

a) There is no physical realization that we know of for the lower-symmetry
fixed-point towards which the system rescales when one of the
symmetry-breaking perturbations is different from zero. All physical
observables are derivatives evaluated at a zero value of the
symmetry-breaking fields $v_s$: In other words, the exponents $\psi _s$ are
crossover exponents associated with the symmetric fixed-point.

b) There is an additional freedom to rescale ${\bf k}$ at each iteration as
seen in Eq.(\ref{scalingG-vs}). This allows one to formulate the
renormalization group in such a way that only a finite number of operators
are relevant! Indeed, for the usual percolation fixed-point, the rescaling
of the operators is found by choosing that the coefficient of the spatial
gradient term in Eq. (\ref{4.15}) to be a constant. Since the recursion
relations for $u_3$ and $r_0$ are completely independent of ${\bf k}$, the
geometric percolation fixed-point is the usual one. The scale factor for $%
{\bf k}$, by contrast, may be chosen at will. This influences the recursion
relations for the $v_s$ and hence the corresponding $\psi _s$ exponents.
More specifically, the rescaling part of the renormalization group
transformation may be written as follows
\begin{equation}
\label{4.49}\Phi _{{\bf k}}\left( q\right) \longrightarrow \Phi _{{\bf k}%
^{\prime }\equiv a{\bf k}}^{\prime }\left( q^{\prime }\equiv qb\right)
=b^{\left( -d-2+\eta _p\right) /2}\Phi _{{\bf k}}\left( q\right) .
\end{equation}
As an example, we choose the scale factor $a$ such that the total resistance
is kept constant under rescaling of all lengths by a factor $b$. This is
done by first noting that after eliminating the degrees of freedom, the
scaling of $v_s$ is obtained by keeping the corresponding terms of the
action invariant
\begin{equation}
\label{4.50}v_s^{\prime }b^{g^{*}(2+d_s)}k^{\prime \,2s}\left[ \Phi _{{\bf k}%
^{\prime }}^{\prime }\left( q^{\prime }\right) \right] ^2\left( \Delta
k^{\prime }\right) ^{MN}d^dq^{\prime }=v_sk^{2s}\left[ \Phi _{{\bf k}}\left(
q\right) \right] ^2\left( \Delta k\right) ^{MN}d^dq.
\end{equation}
In the limit $NM\rightarrow 0$, $(\Delta k)^{NM}$ does not come in since the
infinite-system limit $(\Delta k=0)$ is taken last. Setting $a=b^y$, and $%
b=e^l$, we obtain,
\begin{equation}
\label{4.51}\frac{dv_s}{d\ell }=\left[ 2-\eta _p-g^{*}\left( 2+d_s\right)
+2sy\right] v_s.
\end{equation}
Choosing
\begin{equation}
\label{4.52}y=-\frac 12\left[ 2-\eta _p-g^{*}\left( 2+d_1\right) \right]
\end{equation}
to keep the voltage across the network constant (i.e. $v_1$ does not scale),
the exponents associated with $v_s$ become
\begin{equation}
\label{4.53}\psi _s\rightarrow \psi _s-2s\,\psi _1
\end{equation}
so that the fields $v_s$ now appear irrelevant for $s\geq 2$. In other
words, since the $\psi _s$ are a decreasing function of $s$, we may always
choose the scaling dimension of ${\bf k}$ such that only a few of the $\psi
_s$ exponents are positive, without influencing the Physics. These
statements can be rephrased in a more physical way by recognizing that the
field theory of PHL corresponds to computing the power dissipated between
points $x$ and $x^{\prime }$ when a unit current is injected between these
points, whatever the distance between $x$ and $x^{\prime }$: One could just
as well decide to rescale at unit voltage instead of unit current, and this
would correspond to multiplying ${\bf k}$ by a scale factor at each
iteration. The size dependence of $\Lambda _k$ does not matter, since, in
the limit $\sigma _o^{-1}\rightarrow 0$, $\Lambda _k$ goes to infinity
independently of the system size. The rescaling in ${\bf k}$ is associated
with the scaling of applied voltage so that the scaling in ${\bf k}$ and $%
{\bf q}$ space are independent of each other.

The remark of the last paragraph may also be formulated as follows: The
analysis that we have done to find the eigenpolynomials for Eq.(\ref{4.19})
shows that the latter equation is like a ''second renormalization group''
with fields $v_s$ which describe the electrical properties of an object
whose (critical) geometrical properties are given by the ''first
renormalization group'', with fields $r_0$ and $u_3.$ The first
renormalization group has properties totally independent from those of the
second while the second is slaved to the first. The $v_s$ are conjugate to
replica-space gradients of the field operators $\Phi .$ In other words, they
are conjugate to polynomials of degree $2s$ in ${\bf k}$ times $\Phi _{{\bf k%
}}^2.$ The rescaling of ${\bf k}$ in the second renormalization group is
arbitrary, and this is fundamentally due to the linearity of Kirchhoff's
laws. Instead of referring to relevant or irrelevant exponents, for that
second renormalization group it makes more sense to call them ''dominant''
exponents since one expects that the observables which are connected to $%
\psi _s$ are also coupled to other operators giving corrections to scaling
(sub-dominant exponents).

\subsection{Observability}

The $\varphi _n$ of the $XY$ model are not all relevant exponents. In fact,
for $n\geq 4$, they correspond to irrelevant operators.\cite{mf}\cite
{fourcadexy}\cite{xy}\cite{aharony} While they are only a subset of all
possible irrelevant exponents, they are, however, special because, for
increasing values of $n$, they represent the leading scaling behavior of
operators with lower and lower symmetry. It is their symmetry instead of
their relevance which seems fundamental for their observability. The same
remark applies for the ''dominant'' exponents $\psi _s/\nu \equiv -x_n$
discussed above.

\section{Conclusion}

We have shown that the field theory for multifractals in percolation has a
special structure which allows multifractal exponents to have properties
which do not usually appear in standard critical phenomena. They follow from
symmetry-breaking operators in a ''second renormalization group'' to which
an additional normalization freedom (e.g. scaling at constant voltage or
constant current) is associated. This freedom allows one to arbitrarily
shift the crossover exponents (while maintaining the observable quantities
unchanged). We propose to call these exponents dominant since, even though
their value can be shifted, they are trivially related to the leading
scaling behavior of operators characterized by a given symmetry. A similar
structure with a ''second renormalization group'' also occurs in dynamical
systems\cite{cercle}, and probably also in the field of localization.

\section{Acknowledgments}

A.-M.S.T. acknowledges the support of the Natural Sciences and Engineering
Research Council of Canada (NSERC), the Fonds pour la Formation de
Chercheurs et l'Aide \`a la Recherche (FCAR) of the Government of Qu\'ebec,
the Canadian Institute of Advanced Research (CIAR) and the Killam
Foundation. B. F. acknowledges the friendly hospitality of the Physics
department of the Universit\'e de Sherbrooke where this work was done.

\appendix

\section{Recursion relation to one-loop}

In this Appendix we derive the R.G. Eq. (\ref{self-Laplace}). Let us recall
that, to one-loop order, the recursion trelation for $r_{{\bf k}}$ takes the
form
\begin{equation}
\label{a1}\frac{dr_{{\bf k}}}{dl}=\,\left( 2-\eta _p\right) r_{{\bf k}%
\,}-g(2+\widetilde{\Sigma }_{{\bf k}})
\end{equation}
where the self-energy is defined by
\begin{equation}
\label{a2}\widetilde{\Sigma }_{{\bf k}}=\lim _{n\rightarrow
0}\,\,\int_{-\infty }^{+\infty }\cdot \cdot \cdot \int_{-\infty }^{+\infty
}dp_{\alpha \beta }\,\left[ G\left( {\bf p+k}\right) G\left( {\bf p}\right)
\right]
\end{equation}
with
\begin{equation}
\label{a3}\left[ G\left( {\bf k}\right) \right] ^{-1}=1+p-p_c+\sum_{s\geq
1}v_sk^{2s}P_{\left( s\right) }^{(1)}\left( \cdot \cdot \cdot ,\theta _{{\bf %
k}},\cdot \cdot \cdot \right) .
\end{equation}
Eq. (\ref{a1}) is a non-linear recursion relation for the $v_s$. We now show
how one can linearize this recursion relation in the limit where $%
n\rightarrow 0$ and $\sigma _0^{-1}\rightarrow 0$. Using the Schwinger
representation for the propagator

\begin{equation}
\label{a}G\left( {\bf k}\right) =\int_0^{+\infty }du\text{ }\exp \left[
-u\left( G\left( {\bf k}\right) \right) ^{-1} \right]
\end{equation}
we get for Eq. (\ref{a2})

\begin{equation}
\label{a5}\lim _{n\rightarrow 0}\int_0^{+\infty }du\int_0^{+\infty
}dt\int_{-\infty }^{+\infty }\cdot \cdot \cdot \int_{-\infty }^{+\infty
}dx_{\alpha \beta }\exp \left\{ -u\left[ G\left( {\bf x}\right) \right]
^{-1}-t\left[ G\left( {\bf x+k}\right) \right] ^{-1}\right\} .
\end{equation}
By making the change of variables
\begin{equation}
{\bf x\rightarrow x-}\left[ \frac t{t+u}\right] {\bf k}
\end{equation}
Eq. (\ref{a5}) can be expressed as

\begin{equation}
\label{a6}
\begin{array}{c}
\lim _{n\rightarrow 0}\int_0^{+\infty }du\int_0^{+\infty }dt\int_{-\infty
}^{+\infty }\cdot \cdot \cdot \int_{-\infty }^{+\infty }dx_{\alpha \beta }
\\
\exp \left[ -uG\left[ {\bf x-}\left( \frac t{t+u}\right) {\bf k}\right]
^{-1}-tG\left[ {\bf x+}\left( \frac u{t+u}\right) {\bf k}\right]
^{-1}\right] .
\end{array}
\end{equation}

For $v_s << v_1$, we can expand to first order in $v_s$ all terms in $G$
which are not invariant under rotation (the only term which is invariant is $%
{\bf k}^2$ and it is associated with the resistance). It is useful to define

\begin{equation}
\label{ajout}P\,\equiv \,\sum_{s\geq 2}v_s{\bf k}^{2s}P_s(...,\theta _{{\bf k%
}},...)
\end{equation}
Therefore, Eq. (\ref{a6}) reads as

\begin{equation}
\label{a7}
\begin{array}{c}
-\lim _{n\rightarrow 0}\int_0^{+\infty }du\int_0^{+\infty }dt\int_{-\infty
}^{+\infty }\cdot \cdot \cdot \int_{-\infty }^{+\infty }dx_{\alpha \beta }
\\
\exp \left[ -\left( u+t\right) \left( 1+p-p_c\right) -v_1
{\bf x}^2\left( u+t\right) -v_1{\frac{tu}{t+u}}{\bf k}^2\right] \times \\
u\left[ P\left[ {\bf x-}\left( \frac t{t+u}\right) {\bf k}\right] +P\left[
{\bf x+}\left( \frac t{t+u}\right) {\bf k}\right] \right]
\end{array}
\end{equation}
To interpret Eq. (\ref{a7}) observe that in dimension $n$
\begin{equation}
\begin{array}{c}
\int_{-\infty }^{+\infty }dx_1...\int_{-\infty }^{+\infty }dx_n\,exp
\big[ -v_1(u+t)(x_1^2+...+x_n^2)\big] x_{\beta _1}^{2p_1}...x_{\beta
_j}^{2p_j} \\ = \\
\big( {\frac \pi {v_1(u+t)}}\big)^{\frac n2}\prod_{i=1}^{i=j}\Big[ {\frac{%
(2p_i)!}{2^{p_i}p_i!}}\Big[ {\frac 1{2v_1(u+t)}}\Big]^{p_i}\Big]
\end{array}
\end{equation}
which gives as $n\rightarrow 0$
\begin{equation}
\prod_{i=1}^{i=j}\Big[ {\frac{(2p_i)!}{2^{p_i}p_i!}}\Big[ {\frac 1{2v_1(u+t)}%
}\Big]^{p_i}\Big]
\end{equation}
For analytic functions $P$, we can write with ${\bf k}^{\prime }={\frac t{u+t%
}}{\bf k}$
\begin{equation}
\begin{array}{c}
P(
{\bf k}^{\prime }-{\bf x})+P({\bf x}+{\bf k}^{\prime })= \\ 2\,\sum_{l\geq
0}\sum_{p_1+...+p_j=l}\,{\frac 1{(2p_1)!\,...(2p_j)!}}%
\,x_{_1}^{2p_1}...x_{_j}^{2p_j}{\frac{\partial ^{2p_1+...+2p_j}P}{\partial
k_{_1}^{,2p_1}...\partial k_{_j}^{,2p_j}}}
\end{array}
{}.
\end{equation}
The integrals can then be evaluated as
\begin{equation}
\begin{array}{c}
\lim _{n\rightarrow 0}\int_{-\infty }^{+\infty }d^n
{\bf x}\exp [-v_1(u+t){\bf x}^2]\,\big[P({\bf x}-{\bf k}^{\prime })+P({\bf x}%
+{\bf k}^{\prime })\big]= \\ 2\,\sum_{l\geq 0}\,
{\frac 1{(4v_1(u+t))^l}}\,\sum_{p_1+...+p_j=l}\,{\frac 1{p_1!...p_j!}}\,{%
\frac{\partial ^{2p_1+...+2p_j}P}{\partial k_{_1}^{,2p_1}...\partial
k_{_j}^{,2p_j}}=} \\ 2\exp \big[ {\frac{{\bf \Delta }_{{\bf k}^{\prime }}}{%
4v_1(u+t)}}\big]\,P({\bf k}^{\prime })=2\exp \big[{\frac{u+t}{4v_1}}t^{-2}%
{\bf \Delta }\big] P({\bf k})
\end{array}
\end{equation}
where ${\bf \Delta }$ is the Laplacian operator for the variable ${\bf k}$.
In this equation, we have used the fact that for s$\geq 2$ the term $v_1{%
\frac{tu}{t+u}}k^2$ in Eq.(\ref{a7}) cannot play any role to linear order in
the $v_s$. Now that we have interpreted Eq.($\ref{a7}$), we can substitute
it in Eqs.(\ref{a6}), (\ref{a5}) and (\ref{a2}) to obtain,

\begin{equation}
\label{a8}
\begin{array}{c}
\,\lim _{\sigma _o^{-1}\rightarrow 0}\lim _{n\rightarrow 0}\,
\widetilde{\Sigma }_{{\bf k}}= \\ \,-\lim _{\sigma _o^{-1}\rightarrow
0}\,\sum_{s\geq 2}2v_s\,\int_0^\infty du\int_0^\infty dt\,exp
\big[ -(u+t)(1+r_0)\big] \Big[ {\frac t{u+t}}\Big]^{2s}u \\ \left[ \exp
\left[ \frac{u+t}{4v_1}t^{-2}\,{\bf \Delta }\right] P_s\left( {\bf k}\right)
\right] .
\end{array}
\end{equation}
Because the thermodynamic limit is taken last in replica approaches, we can
show, as is done in the text, that $\exp \left[ \frac{u+t}{4v_1}t^{-2}\,{\bf %
\Delta }\right] $ has a well-defined limit as $\sigma _o^{-1}\rightarrow 0$.

The case s=1 can now be treated simply. Indeed, in this case, it is only the
term $v_1{\frac{tu}{t+u}}k^2$ which plays a role. Expanding it, the term
linear in $v_1$ in $\widetilde{\Sigma }_{{\bf k}}$ is given by%
$$
\lim _{\sigma _o^{-1}\rightarrow 0}v_1\int_0^\infty du\int_0^\infty dt\exp
\left[ -\left( u+t\right) \right] \frac{tu}{t+u}{\bf k}^2.
$$
This shows that ${\bf k}^2$ is an eigenpolynomial, as quoted in the text.
Furthermore, the corresponding eigenvalue $d_1$ does correspond to the s=1
limit of Eq.(\ref{4.43}).

\end{document}